\documentclass[10pt, oneside, twocolumn]{article}

\usepackage[a4paper, 
left=1.5cm, right=1.5cm, top=1.5cm, bottom=2.5cm,
]{geometry}
\usepackage{newtxtext}
\usepackage{newtxmath}
\usepackage{microtype}
\usepackage[version=4]{mhchem}
\usepackage{siunitx}
\usepackage{amsmath}
\usepackage{booktabs}
\usepackage{xcolor}
\usepackage[normalem]{ulem}

\usepackage{graphicx}
\graphicspath{{fig/}}

\usepackage[
    style=numeric-comp,
    autocite=superscript, 
    sorting=none, 
    url=false, 
    hyperref=auto, 
    giveninits=true, 
    maxnames=3, 
    minnames=1, 
]{biblatex}
\usepackage{hyperref}
\hypersetup{
    colorlinks=true,  
    linkcolor=blue,   
    citecolor=blue,   
    urlcolor=blue     
}
\usepackage{url}
\newcommand{\XS}{\mathrm{XS}}

\title{Uncertainties in interfacial excess calculations from atom probe tomography data}
\author{Levi Tegg and Julie M. Cairney$^*$}
\date{\small{
  School of Aerospace, Mechanical and Mechatronic Engineering, and\\
  Australian Centre for Microscopy and Microanalysis\\
  The University of Sydney, Camperdown, NSW 2006, Australia\\
  $^{*}$ Corresponding author:  \href{mailto:Julie.Cairney@sydney.edu.au}{Julie.Cairney@sydney.edu.au}
}}

\begin{document}

\maketitle


\section*{Abstract} 
Atom probe tomography can be used to measure the excess of solute chemical species at internal interfaces, but common protocols do not usually consider the compositional uncertainty of the (usually dilute) solute.
Here, general models are derived for the composition profile of a solute segregated to the interface between two grains or phases.
This model is fit to experimental data, with the cumulative compositional uncertainty included in the calculation of the interfacial excess.
It is shown that the relative uncertainty in the interfacial excess can be large for dilute solutes, even when segregation is obvious in the concentration profile.
Different methods for estimating the extent of the boundary provide different estimates for the interfacial excess and its uncertainty.
Some strategies are provided for the minimisation of uncertainties in typical data, though ultimately, a typical calculation will be limited by compositional uncertainty arising from the relatively few solute counts in the region-of-interest.


\section{Introduction}
Atom probe tomography (APT) is a valuable tool for studying the composition of interfaces in solid materials \autocite{GaultTextbook_2012, Miller_Forbes_2014, 
}.
Examples include solute segregation to grain boundaries \autocite{Hudson_Smith_2009, Felfer_2012, Li_Ponge_Choi_Raabe_2015} or phase boundaries \autocite{Yoon_2004, Tytko_2012, Bagot_Reed_2017} in alloys, the internal interfaces of nanoporous metals \autocite{El-Zoka_2017}, or grain boundaries in photovoltaic semiconductors \autocite{Couzinie_2015, Schwarz_2018}.
To perform such an analysis, the atom probe data is reconstructed to produce a three-dimensional map of the collected ions, and composition profiles are calculated along a suitably-positioned region-of-interest (ROI) \autocite{Krakauer_Seidman_1993}, or from a proximity histogram (proxigram) of an isoconcentration surface (isosurface) \autocite{Hellman_Seidman_2002, Yoon_2004}.
In either case, differences in the rate of field evaporation across the specimen can distort the reconstructed volume and thus the measured composition profile.
For example, differences in evaporation field between phases can lead to local magnification \autocite{Miller_Hetherington_1991}, distorting the shape of precipitates and subsequent composition profiles.
To compensate, interfacial excess calculations can be performed in a way that the reconstructed ionic density does not affect the calculated composition \autocite{Krakauer_Seidman_1993}.
These calculations give the number of solute ions per unit area on some defined interface.
The interfacial excess is usually determined graphically from a plot of the cumulative count of the solute ions against the cumulative count of all ions, both of which are found from a composition profile 
\autocite{Krakauer_Seidman_1993, Li_Ponge_Choi_Raabe_2015, El-Zoka_2017, Langelier_2017, Schwarz_2018, Jenkins_2020, Theska_Primig_2024}.
Such a graph is sometimes known as a ``ladder plot'' \autocite{Babu_1994}.

Despite the maturity of interfacial excess calculation protocols and their wide use in the community, reporting of uncertainties is limited.
Many articles report interfacial excesses without any uncertainties, even if the uncertainties in the composition profile are available \autocite{Miller_Smith_1995, Hellman_Seidman_2001, Yoon_2004, Li_Ponge_Choi_Raabe_2015, El-Zoka_2017, Schwarz_2018}.
Others report uncertainties but do not indicate how they were calculated \autocite{Tytko_2012, Bagot_Reed_2017}.
Relatively few studies have published uncertainties and their methods for doing so.
In the original report of an interfacial excess calculation using APT, Krakauer and Seidman \autocite{Krakauer_Seidman_1993} determined uncertainty in the interfacial excess from the uncertainty in the composition and in the area of their detector.
Using the compositional uncertainty (in some form) is the most common way to estimate uncertainties in interfacial excess calculations.
Another approach is to produce multiple smaller ROIs, calculate the interfacial excess of each, and then estimate the uncertainty of the average value using some statistical metric, like the standard deviation \autocite{Hudson_Smith_2009, Jiang_Rouxel_Langan_Dorin_2021}.
The shortcoming of this method is that it trades the high statistical power of one large profile (with many ions) for the poor statistical power from many small profiles (each containing far fewer ions).

The shortcomings of the typical interfacial excess calculation have been recognised by the community, and there have been reports on the optimisation and automation of interfacial excess calculation protocols \autocite{Felfer_2015, Peng_2019, Blum_2020, Jenkins_2020, Theska_Primig_2024}.
Jenkins et al. identified that the extent of the interface is subjective and the excess calculated depends on the definition of that extent \autocite{Jenkins_2020}.
Felfer et al. studied analysis of curved surfaces using Voronoi tessellation \autocite{Felfer_2013}, setting the polygon size according to the required solute counts for a particular statistical confidence \autocite{Felfer_2015}.
A similar though more complex technique was reported by Peng et al. \autocite{Peng_2019}.
Blum et al. \autocite{Blum_2020} used an ``integral'' method \autocite{Huang_2016} which considers uncertainties and the potential for smooth variation in solute content between grains, but performs the analysis with a position abscissa, rather than total ion count.
A similar method was reported by Maugis and Hoummada \autocite{Maugis_Hoummada_2016}, though their method neglects uncertainties and may not apply to systems with different solute concentration on different sides of an interface.
Theska and Primig \autocite{Theska_Primig_2024} considered the uncertainty in the solute counts in each grain, but did not describe how the extent of the interface was defined.

Here, we describe a curve-fitting method to include compositional uncertainty into interfacial excess calculations.
General e`xpressions are derived for the concentration profile of a segregated solute at an interface.
These expressions are fit to experimental compositional profiles, and parameters derived from the models give the interfacial excess and its uncertainty.
Importantly, the uncertainty in the cumulative solute count is the sum of the compositional uncertainty in every previous step.
We show that the uncertainty in the interfacial excess is quite large for dilute solutes, even when the segregation is obvious from the composition profile.
This work concludes with some recommendations for minimising uncertainty in interfacial excess calculations, and methods for its accurate calculation.

\section{Derivation and modelling}
\begin{figure}[t!]
  \centering
  \includegraphics[width=\linewidth]{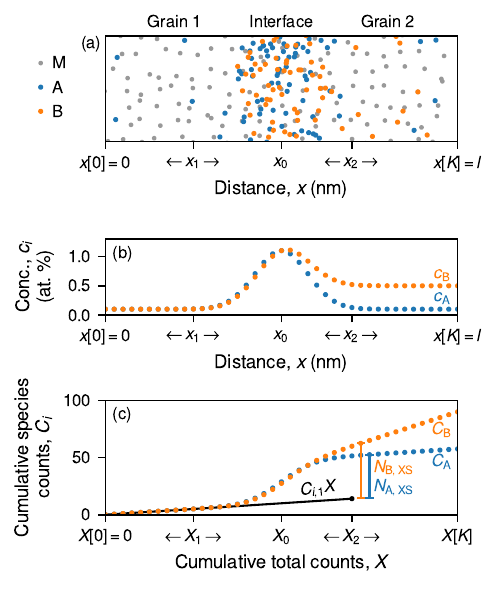}
  \caption{(a)~Model atom probe data showing solutes ($i = \mathrm{A, B}$) segregated to the interface between two grains of a matrix species (M). (b)~Model concentration profile over this interface for each solute, based on equations \ref{eq:conc_equal} and \ref{eq:conc_unequal}. (c)~Model cumulative count profile over this interface for each solute, based on equations \ref{eq:cumcount_equal} and \ref{eq:cumcount_unequal}. Annotations indicate the extrapolation of the linear region ($C_{i,1}X$) and calculations of $N_{i,\XS}$ at $X_2$ using equations \ref{eq:NXS_equal} and \ref{eq:NXS_unequal}.}
  \label{fig:schematic}
\end{figure}

An interfacial excess calculation quantifies the excess or depletion of a chemical species $i$ at an interface, above or below its concentration elsewhere in the material.
This section will use a model system with segregation of two solutes to the interface between two grains of some matrix, in which the solutes have low but non-zero concentration.
This is illustrated in figure \ref{fig:schematic}(a) for matrix species M, and solute species $i = \mathrm{A}$ and B.

One common method for calculating the interfacial excess is to analyse the plot of the cumulative counts of the ion of interest ($C_i$) against the cumulative counts of all ions ($X$) along some region-of-interest, also known as the ladder plot. 
The cumulative ion count for species $i$ for discrete data is
\begin{equation} \label{eq:cumulativesum}
  C_i[k] = \sum_{j = 0}^{k} N[j] c_i[j] \,.
\end{equation}
Here, $k = 0, 1, 2, ..., K$ and $j = 0, 1, 2, ..., k$ are integer indices for the vectors $x$, $N$ and $c_i$, which are respectively the position along the region-of-interest (of length $l$), the total counts at position $x$, and the concentration of species $i$ at $x$.
The concentration is found from $c_i = N_i/N$, where $N_i$ is the counts of $i$ \autocite{GaultTextbook_2012}.
For example, a ROI of $l = 20$~nm and 0.5~nm bin width (and thus $K = 40$) would have $k = 0, 1, 2, ..., 40$, giving $x[k] = $ 0~nm, 0.5~nm, 1.0~nm, $...$, 20 nm.
We use this parametrisation for easy comparison with data produced by Cameca IVAS, and ease of implementation in programming languages where vector indexing starts at zero, such as Python.

Interfacial excess calculations can be performed using $x[k]$ on the abscissa only if $N$ is constant with $x$.
In real data, effects such as local magnification, laser heating, and crystallographic poles lead to uneven rates of evaporation across the specimen, and thus an unequal number of ions per position step.
As such, it is more appropriate to perform interfacial excess calculations using the cumulative count of all ions, $X$,
\begin{equation} \label{eq:cumulativetotalsum}
  X[k] = \sum_{j = 0}^{k} N[j]\,.
\end{equation}
It is worth emphasising that $X$ increases with $x$ and $k$, is always positive, and the length of the vectors $k$, $x$ and $X$ are the same.
These properties allow us to treat $X$ as if it were a function of $x[k]$, and although $x$ and $X$ are not generally equivalent, either could be used as the abscissa in an interfacial excess calculation.
In this work, concentration ($c_i$) profiles will be expressed in terms of position ($x$), and cumulative ion count ($C_i$) profiles will be expressed in terms of cumulative total count ($X$).
Lower case symbols will be used for concentration and position quantities, and upper case for ion count quantities.
Although matching sets of lower and upper case symbols have similar meaning, they do not necessarily have equal value, as explained below.

\subsection{Grains of equal solute concentration} \label{section.equal}

A model for the concentration profile of a segregated solute at an interface is the sum of a constant concentration and a Gaussian function centred at the boundary:
\begin{equation} \label{eq:conc_equal}
 c_i(x) = c_{i,0} + g\mathrm{e}^{-(x-x_0)^2 / (2w^2)}
\,.
\end{equation}
Here, $c_{i,0}$ is the concentration of $i$ in the matrix grains, $g$ is the magnitude of the solute segregation peak, $0 \leq x_0 \leq l$ is the position of the boundary, and $w$ is the width of the peak.
An example of this is shown for solute A in figure \ref{fig:schematic}(b).
Although $x$ and $c_i$ are defined only at $x[k]$ positions, the other parameters in equation \ref{eq:conc_equal} can take any real value.
The cumulative distribution function (CDF) of this profile is
\begin{equation} \label{eq:cumcount_equal}
  C_i(X) = C_{i,0}X + G W \sqrt{\frac{\pi}{2}} \left(1 + \mathrm{erf}\left( \frac{X - X_0}{W \sqrt{2}} \right) \right) \,,
\end{equation}
where $\mathrm{erf}()$ is the error function \autocite{Ng_Geller_1969}, 
$C_{i,0}$ represents the solute concentration in the matrix,
$G$ is the magnitude of the segregation peak, $X_0$ is the position of the boundary, and $W$ is the width of the peak.
An example of this profile is shown with solute A in figure \ref{fig:schematic}(c).

For data where $X[k] = Nk$ and there is constant $N$ with position, upper case parameters are related to their lower case counterparts by $C_{i,0} = c_{i,0}$, $G = g$, $X_0 = x_0 N/(x[k+1]-x[k])$ and $W = wN/(x[k+1]-x[k])$.
If $N[k]$ varies little with $x[k]$, the average $\bar{N}$ over the ROI can be used in place of $N$.
If $N$ varies significantly with $x$, there is no straightforward equivalence between the parameters.

Within the first grain (any $X_1 \ll X_0$), $C_i$ is linear with gradient $C_{i,0}$.
Around the grain boundary ($X \approx X_0$), $C_i$ increases proportional the to the ratio of $G/C_{i,0}$.
In the second grain ($X_2 \gg X_0$), $C_i$ again becomes linear with gradient $C_{i,0}$.

The difference between the $C_i$ from equation \ref{eq:cumcount_equal} and that predicted from the linear region from grain~1,
\begin{equation} \label{eq:cumcount_linear}
  C_i(X) = C_{i,0}X \,,
\end{equation} 
can give the excess \textit{counts} of the solute ions at the interface:
\begin{equation} \label{eq:NXS_equal}
  N_{i,\XS}(X_2) = C_i(X_2) - C_{i,0}X_2 \,.
\end{equation}
This is indicated in figure \ref{fig:schematic}(c).
Note that this is evaluated at $X_2$ (i.e. $k_2$), some position (i.e. index) within the second grain.
When the solute concentration in the matrix is equal in both grains and there is no compositional uncertainty, $X_2$ can be chosen anywhere in the second grain that is sufficiently far from the interface.

$N_{i,\XS}$ is used to calculate the Gibbsian interfacial excess of \textit{atoms} at the interface, $\Gamma_i$:
\begin{equation}
  \Gamma_i = \frac{N_{i,\XS}(X_2)}{\epsilon A} \,,
\end{equation}
which accounts for the detector efficiency $0 \leq \epsilon \leq 1$ and the cross-sectional area of the region-of-interest, $A$.
If the average interatomic spacing along the direction of the region-of-interest is known (for example, if it is parallel with a crystal lattice direction) then the interfacial excess can be expressed in terms of equivalent monolayers of the matrix structure using
\begin{equation}
  \Phi_i = \frac{\Gamma_i}{\rho d}
\end{equation}
where $\rho$ is the volumetric atom number density and $d$ is the interatomic spacing.
Choices of $X_2$ and thus methods for calculating $N_{i, \XS}$ are described later in this work.

\subsection{Grains of unequal solute concentration}

If the grains do not have equal solute concentration then shape of $C_i$ changes, and the choice of $X_2$ is more important.
One way to model this scenario is to use another error function to represent the change in solute content in the grains:
\begin{equation} \label{eq:conc_unequal}
  \begin{split}
  c_i(x) & =
  c_{i,1} \\
  & + g\mathrm{e}^{-(x-x_0)^2 / (2w^2)} \\
  & + \frac{c_{i,2} - c_{i,1} }{2} \left( 1 + \mathrm{erf}\left( \frac{x - x_0}{w \sqrt{2}} \right) \right)  \,,
  \end{split}
\end{equation}
where $c_{i,1}$ and $c_{i,2}$ are the solute concentrations in the first and second grain, and the other symbols have the same meaning as in equation \ref{eq:conc_equal}.
In equation \ref{eq:conc_unequal}, the first term represents the concentration in the first grain, the second term the segregation at the interface, and the third term the smooth change in concentration to the second grain.
This scenario is shown for solute B in figure \ref{fig:schematic}(b).

Although the true shape of the third term is more likely a Heaviside step function, the point spread function in atom probe data is Gaussian \autocite{Gault_Moody_2010}, so the corresponding CDF is more appropriate.
This approach is similar to that used by Blum et al. \autocite{Blum_2020, Huang_2016}, except that an error function is used here instead of a hyperbolic tangent, which does not have an obvious physical basis.
Additionally, the width ($w$) of solute peak (i.e. Gaussian peak) and the gradual change between grains (i.e. error functions) are not likely to be equal in a real material. 
However, atom probe data does not usually have sufficient spatial resolution to determine them separately.
To reduce covariance and uncertainties in the least-squares analysis, this derivation will use the same width parameter for both terms.

The CDF of equation \ref{eq:conc_unequal} is\autocite{Ng_Geller_1969}
\begin{equation} \label{eq:cumcount_unequal}
  \begin{split}
  C_i(X) & = C_{i,1} X \\
  & + \left(G - \frac{C_{i,2} - C_{i,1}}{2} \right) W \sqrt{\frac{\pi}{2}} \left(1 + \mathrm{erf}\left( \frac{X - X_0}{W \sqrt{2}} \right) \right) \\
  & + \frac{C_{i,2} - C_{i,1}}{2} 
     \begin{split} & \left(  X 
      + (X - X_0) \mathrm{erf} \left( \frac{X - X_0}{W \sqrt{2}} \right) \right. \\
      & \left. + W \sqrt{\frac{2}{\pi}} \mathrm{e}^{-(X-X_0)^2 / (2W^2)}  \right) \end{split}    \\
  & - \frac{X_0 (C_{i,2} - C_{i,1})}{2}  \,,
  \end{split}
\end{equation}
where the four terms represent the contributions from grain 1, the  interface, the smooth change to grain 2, and the integration constant found from the boundary condition $C_i(0) = 0$.
An example of this profile is shown for solute B in figure \ref{fig:schematic}(c).

As before, the difference between the $C_i$ from equation \ref{eq:cumcount_unequal} and that predicted from the linear region from grain 1 gives the excess counts of the solute across the interface:
\begin{equation}  \label{eq:NXS_unequal}
  N_{i,\XS}(X_2) = C_i(X_2) - C_{i,1}X_2 \,.
\end{equation}
However, the choice of $X_2$ is now important because the gradients of the two linear sections are not equal. 
$X_2$ must be chosen far enough above $X_0$ such that it does not lie within the interfacial region, but close enough to $X_0$ that the varying difference between $C_{i,1}$ and $C_{i}$ does not influence the result.
Some strategies for choosing $X_2$ will be described in this work.

\subsection{Uncertainties}
If a single, large, representative ROI or isosurface is used to study the interface, then the dominant source of uncertainty in an interfacial excess calculation is the measurement uncertainty for $i$ in each bin of the concentration profile, i.e. the compositional uncertainty.
Finding expressions for the uncertainty in the parameters of equations \ref{eq:cumcount_equal} and \ref{eq:cumcount_unequal} is unnecessary, as these parameters will later be fit using a least-squares analysis and standard deviation estimates will be determined there.
Instead, a brief overview of the measurement uncertainty is more useful here.
Throughout this work, the uncertainty in a quantity ($\Delta$) will be its estimated standard deviation (ESD).

For a dilute solute of low concentration but with counts far greater than the background, the uncertainty in counts is \autocite{Gedcke_2009, Larson_2013}
\begin{equation}
  \Delta N_i = \sqrt{N_i} \,.
\end{equation}
Since the cumulative ion count is a repeated sum, the uncertainty in the cumulative ion count is the sum of these uncertainties:
\begin{equation} \label{eq:cumulativesum_uncertainty}
    \Delta C_i[k] = \sum_{j = 0}^{k} \sqrt{ N_i[j] } \,.
\end{equation}
Note the sum is performed from $j = 0, 1, 2, ..., k$ to give $\Delta C_i$ at $k$.
This is a key difference to previous reports, which have considered only $\Delta C_i[k] = \sqrt{N_i [k]}$, if any uncertainties were considered at all.

The uncertainty in $C_i$ increases with $x[k]$ and with $c_i$.
This adds further pressure to the choice of $X_2$, as greater $\Delta C_i(X_2)$ will increase uncertainty in the measured excess counts, $\Delta N_{i,\XS}$:
\begin{equation}
  \Delta N_{i,\XS} = \Delta C_i(X_2) + \Delta C_{i,0}(X_2)
\end{equation}
where $\Delta C_{i,0}$ is the uncertainty in the gradient of the linear region from grain 1, which is evalulated at $X_2$.

The relative uncertainties $\Delta N_i/N_i$ and $\Delta c_i / c_i$ are also useful for this work.
Equivalent expressions for the uncertainty in concentration are \autocite{Larson_2013}
\begin{equation} \label{eq:concentration_uncertainty}
  \Delta c_i = c_i \frac{\Delta N_i}{N_i} = \frac{c_i}{\sqrt{N_i}} = \frac{\sqrt{N_i}}{N}
\end{equation}
The first term shows that the relative uncertainties in $c_i$ and $N_i$ are equal.

\subsection{Least-squares fitting}
Equations \ref{eq:cumcount_equal} and \ref{eq:cumcount_unequal} can be fit to cumulative count profiles calculated from experimental concentration profiles.
The ESD in the model parameters can be used to estimate $\Delta C_{i,1}(X_2)$, or $N_{i,\XS}$ directly.
Least-squares fitting of experimental data was performed using the Levenberg-Marquardt algorithm as implemented within the \texttt{optimize.curve\_fit} function within the \texttt{scipy} library \autocite{SciPy_2020} for Python~3.8.
Uncertainties in the experimental $C_i$ were calculated from equation \ref{eq:cumulativesum_uncertainty}.
ESDs of the model parameters were found from the square-root of the diagonal elements of the covariance matrix.
The model parameters were mostly independent, except for some small positive correlation between $X_0$ and $W$.
These parameters do not significantly affect the result of equations \ref{eq:NXS_equal} or \ref{eq:NXS_unequal}, so this correlation was considered acceptable.

\section{Experiment}
The materials studied in this work are Optimized ZIRLO, a Zr-Nb-Sn alloy, and a microalloyed martensitic steel, both described elsewhere \autocite{Huang_2023, Lin_2020}.
Matchstick-shaped sections were sharpened into needles using a two-stage electropolishing procedure with a perchloric acid electrolyte \autocite{Hopkins_1965, GaultTextbook_2012}. 
The Optimized ZIRLO was further sharpened by annular-milling with a Thermofisher Helios G4 Xe plasma focused ion beam to position a grain boundary near the specimen apex \autocite{Huang_2023}.

The Optimized ZIRLO specimen was analysed in a Cameca LEAP4000 X Si atom probe in laser-pulsed mode with a 355 nm laser wavelength, 100 pJ laser pulse energy, 200 kHz pulse repetition rate, 0.5\% target detection rate, and 50 K base temperature \autocite{Huang_2023}. 
This instrument has $\epsilon = 0.57$ detection efficiency.
The steel specimen was analysed in a Cameca Invizo 6000 atom probe in laser-pulsed mode with a 257.5 nm laser wavelength, 400 pJ laser pulse energy, 200 kHz pulse repetition rate, 0.5\% target detection rate, and 50 K base temperature.
This instrument has $\epsilon = 0.62$ detection efficiency \autocite{Tegg_2023}.
Data were reconstructed and analysed using the IVAS module within Cameca AP Suite 6.3. 
The image compression factor and field factor for the steel specimen were determined through analysis of crystallographic poles.
No poles were observed for the Optimized ZIRLO specimen, so the instrument-default reconstruction parameters were used and thus the reconstruction has not been spatially calibrated.
Electric fields were calculated using the Zr +++/++ and Fe ++/+ charge state ratios in the Optimized ZIRLO specimen and the steel specimen, respectively \autocite{Tegg_Stephenson_Cairney_2024}.

Concentration profiles over a grain boundary in the Optimized ZIRLO specimen were calculated using a cylindrical ROI of dimension $\approx 70 \times 100 \times 25$~nm and bin width 0.5 nm.
The cross-sectional area of this ROI is $A \approx 5500\ \mathrm{nm}^2$.
Concentration profiles over a phase boundary between a large cementite (\ce{M3C}) precipitate and the martensite matrix were calculated using a proxigram of bin distance 0.5~nm from an isosurface of the decomposed C concentration.
The isovalue was chosen as $c_{\ce{C}} = 16$ at.\% to position the isosurface near the maximum concentration of the \ce{Mn} profile.
The area of this isosurface is $A \approx 1300\ \mathrm{nm}^2$.

\section{Results}
\subsection{Mass spectra}

\begin{figure}[b!]
  \centering
  \includegraphics[width=\linewidth]{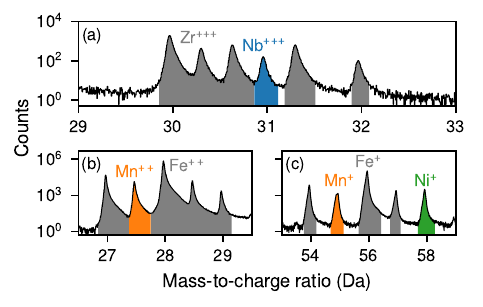}
  \caption{The mass-to-charge ratio spectra (mass spectra) for (a) the Optimized ZIRLO data and (b,c) the steel data. Shaded regions indicate the ranges of the matrix element (grey) and the solutes of interest, either Nb (blue) or Mn (orange). $^{58}\ce{Ni}^+$ overlaps with $^{58}\ce{Fe}^+$ and has been coloured green.}
  \label{fig:ms}
\end{figure}

Figure \ref{fig:ms} shows selected regions of the mass-to-charge ratio spectra (mass spectra) for the two materials studied in this work: (a) the Optimized ZIRLO and (b,c) the steel.
The ranges for the matrix elements (Zr, Fe) are shown in grey, and the ranges for the solutes of interest are shown in (Nb) blue and (Mn) orange.
The solute peaks are prominent, relatively sharp, of far greater magnitude the background, and do not overlap with other species.
There is overlap between $^{58}\ce{Ni}^+$ and $^{58}\ce{Fe}^+$, and the natural abundance of the isotopes of each show this peak is mostly \ce{Ni}. 
Both reconstructions contain more elements and charge states than shown \autocite{Huang_2023, Lin_2020}, though they are not relevant for the analysis described here.

\subsection{Model fit to experimental data}

\begin{figure}[b!]
  \centering
  \includegraphics[width=\linewidth]{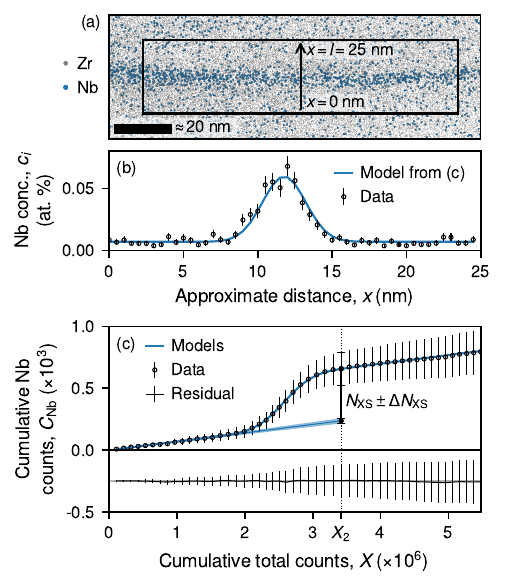}
  \caption{Application of equation \ref{eq:cumcount_equal} to experimental data where the solute has equal concentration on either side of the interface. (a) Reconstructed atom probe data showing Nb segregated to a boundary between $\alpha$-Zr grains \autocite{Huang_2023}. Annotations indicate the dimensions and orientation of a cylindrical ROI. (b) A concentration profile over the ROI showing \ce{Nb} concentration and a model profile fit. (c) Cumulative Nb and total counts from the concentration profile. Data has been fit using equation \ref{eq:cumcount_equal}, and the model is also plotted with in (b) with using the parameter transforms described in section \ref{section.equal}.}
  \label{fig:data_01}
\end{figure}

Figure \ref{fig:data_01}(a) shows a grain boundary in the reconstructed volume from the Optimized ZIRLO specimen with the boundary and orientation of a cylindrical ROI overlaid.
Zr positions are shown in grey and Nb positions in blue.
\ce{Fe}, \ce{Sn} and \ce{Nb} are segregated to this boundary \autocite{Huang_2023}, with the latter chosen for this analysis.
Figure \ref{fig:data_01}(b) shows the concentration profile along the ROI (viewed from the side so that it appears rectangular), and (c) shows the cumulative Nb counts against the cumulative total counts.
Because the Nb concentration is equal in each grain, equation \ref{eq:cumcount_equal} was fit to the data in (c).
The data and model fit are shown for positive $C_i$, with the residual (data minus model) shown for negative $C_i$.
The Nb concentration in each grain is $c_{\ce{Nb},0} = 0.0056 \pm 0.0004$~at.\%, with this uncertainty the ESD in the $c_{i,0}$ model parameter.
The electric field was measured as $\approx 23.5$~V/nm in each grain, with a slight peak to $\approx 24.5$~V/nm in the boundary.
Despite this, the ionic density was relatively uniform over this ROI, so the parameters fit using equation~\ref{eq:cumcount_equal} for the $C_{\ce{Nb}}$ data could be applied to the $c_{\ce{Nb}}$ data using the upper-lower case transformations described in section \ref{section.equal}.
The error bars in (b) are those reported by IVAS and are equal to those found from equation \ref{eq:concentration_uncertainty}.
The error bars in (c) are calculated using equation \ref{eq:cumulativesum_uncertainty}.
Agreement between the data and the model is excellent, both for the cumulative ion count profile and for the composition profile.
An annotation in (c) indicates a choice for $X_2$ and a calculation of $N_\XS$, though this is considered more rigorously later in this manuscript.

\begin{figure}[b!]
  \centering
  \includegraphics[width=\linewidth]{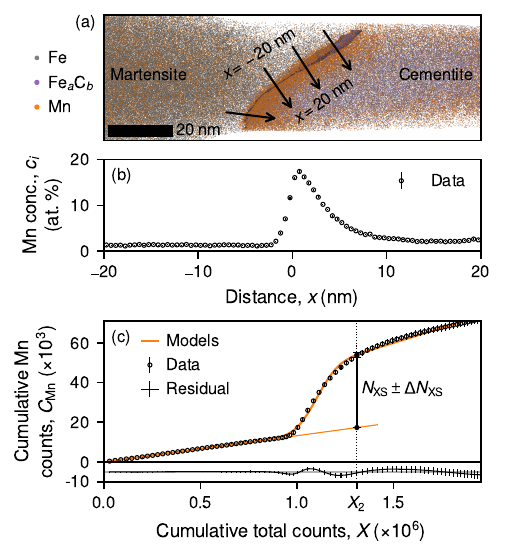}
  \caption{Application of equation \ref{eq:cumcount_unequal} to experimental data where the solute has different concentration on either side of the interface. (a) Reconstructed atom probe data showing \ce{Fe}, \ce{Fe}-\ce{C} complex ions, and \ce{Mn} positions from a sample of the microalloyed steel \autocite{Lin_2020}. The left of the image shows a martensite grain and the right side a cementite (\ce{M3C}) precipitate. An isosurface of decomposed $c_{\ce{C}} = 16$~at.\% is visible. (b) A proxigram over the isosurface showing \ce{Mn} concentration. (c) Cumulative \ce{Mn} and total counts from the proxigram in (b). Data has been fit using equation \ref{eq:cumcount_unequal}.}
  \label{fig:data_02}
\end{figure}

Figure \ref{fig:data_02}(a) shows a phase boundary between a martensite grain (left) and a cementite precipitate (right) in the reconstructed volume from the steel specimen.
An isosurface of C at $c_{\mathrm{C}} = 16$\% was used to calculate a proxigram, with the arrows indicating the direction of increasing $x$.
\ce{Mo}, \ce{Cr} and \ce{Mn} are segregated to the boundary, with the latter chosen for this analysis.
Figure \ref{fig:data_02}(b) shows the \ce{Mn} proxigram over the isosurface and (c) shows the cumulative Mn counts against the cumulative total counts.
Equation~\ref{eq:cumcount_unequal} was fit to the data in (c).
The data and fit are shown for positive $C_i$, with the residual (data minus model) shown for negative $C_i$.
The Mn concentration is $c_{\ce{Mn},1} = 1.29 \pm 0.01$~at.\% in the martensite and $c_{\ce{Mn},2} = 2.9 \pm 0.1$~at.\% in the cementite, again with these uncertainties representing the ESDs of the relevant model parameters.
The electric field was measured as $\approx 19$~V/nm in the martensite and $\approx 20$~V/nm in the cementite with no clear increase or decrease at the interface.
The rate of field evaporation was not uniform over this ROI, so the parameters fit using equation \ref{eq:cumcount_equal} for the $C_{\ce{Nb}}$ data do not describe the $c_{\ce{Nb}}$ when used in equation \ref{eq:conc_unequal}.
The variation in the residual indicates that equation \ref{eq:cumcount_unequal} is not a perfect description of this CDF.
This is clear from figure \ref{fig:data_02}(b), which shows a skewed $c_i$ peak.
Despite the inaccuracies of the model, calculation of $N_{i, \XS}$ can be calculated without sensitivity to the shape of the fitted $C_i$ around $X_0$, so the interfacial excess can still be calculated

\begin{figure}[b!]
  \centering
  \includegraphics[width=\linewidth]{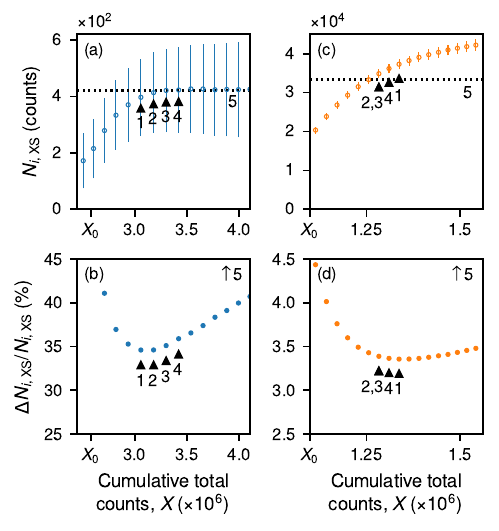}
  \caption{The interfacial excess for (a,b) the Optimized ZIRLO specimen and (b,d) the steel sample. (a,c) Variation in $N_{\XS}$ with $X$. (b,d) Relative uncertainty $\Delta N_{\XS}/N_{\XS}$ in terms of $X$. Markers indicate choices for $X_2$ based on methods 1 through 4, whilst a dashsed line in (a,c) indicates the $N_{i,\XS}$ for method 5. In (b,d), methods 2 and 3 produce the same $X_2$. The relative for uncertainty for method 5 is much greater than the range of the plot. The exact values for each $N_{i,\XS}$ are given in table \ref{table:NXS}.}
  \label{fig:data_NXS}
\end{figure}

Figure \ref{fig:data_NXS} shows the interfacial excess $N_{i, \XS}$ for (a) Nb in the Optimized ZIRLO reconstruction and (c) Mn in the steel specimen, expressed in terms of $X$.
Although $N_{\ce{Nb},\XS}$ is relatively uniform within grain 2, the $\Delta N_{\ce{Nb},\XS}$ continuously increases.
This can be seen in the relative uncertainty $\Delta N_{\ce{Nb},\XS}/N_{\ce{Nb}, \XS}$, shown in figure \ref{fig:data_NXS}(b).
As expected, the $N_{\ce{Mn},\XS}$ continuously increases with $X$ because $c_{i,1} \neq c_{i,2}$.
These results emphasise the importance of a good strategy for choosing $X_2$, as described below.

\subsection{Choice of $X_2$}
It has been identified that the choice of $X_2$ is important when analysing real data \autocite{Jenkins_2020}, and that there are several strategies for doing so.
Here, some methods for choosing $X_2$ are compared.
Triangular markers in figure \ref{fig:data_NXS} indicate the $X_2$ provided by the methods listed below.
The $N_{i,\XS}$ for each method are listed in table \ref{table:NXS}.

\begin{table}[t!] 
  \caption{Values for $N_{i,\XS}$ calculated using the different methods described here. Parenthetical percentages indicate the relative uncertainty.}
  \label{table:NXS}
  \medskip
  \centering
  \begin{tabular}{@{}lll@{}}
    \toprule
    Method              & $N_{\ce{Nb},\XS}$ & $N_{\ce{Mn},\XS}$ \\ \midrule
    1: MRU             & $397 \pm 137$ (35\%) & $37297 \pm 1252$ (3.5\%) \\
    2: $2W$            & $413 \pm 143$ (35\%) & $34863 \pm 1181$ (3.6\%) \\ 
    3: $2\mathrm{HW}$  & $421 \pm 148$ (35\%) & $34863 \pm 1181$ (3.6\%) \\ 
    4: 99\%ERF     & $423 \pm 152$ (36\%) & $36192 \pm 1218$ (3.5\%) \\
    5: Param & $421 \pm 194$ (46\%) & $33368 \pm 2157$ (6.5\%) \\    \bottomrule
    \end{tabular}
\end{table}

Method 1: the minimum relative uncertainty (MRU) of $N_{\XS}$.
The position of the MRU can be interpreted as a threshold between the interfacial and linear regions, as $N_{i,\XS}$ increases more quickly than $\Delta N_{i,\XS}$ near the interface, but much more slowly in grain 2.
An advantage of this method is that it does not require the fitted $C_i$ accurately describe the experimental data, it only requires the $\Delta c_i$ vary little with $x$.
Method 1 produces the lowest $X_2$ in the Optimized ZIRLO reconstruction but the highest $X_2$ in the steel reconstruction.

Method 2: two standard deviations (i.e. $2W$) above $X_0$.
For a Gaussian distribution (as used here) this is equivalent to the 95\% confidence interval, a common statistical metric \autocite{Felfer_2015}.
Note that for the $N_{\ce{Mn},\XS}$ data, methods 2 and 3 give equal $X_2$.

Method 3: two half-width at half-maxima (i.e. $2\mathrm{HW}$) above $X_0$. 
The popularity of the full-width half-maximum measure in microscopy inspired this method.
Since $\mathrm{HWHM} = \sqrt{2 \mathrm{ln}(2)} W \approx 1.177W$, this method may produce the same $X_2$ as the $2W$ method for data with coarse bin widths.  
This is the case for the $N_{\ce{Mn},\XS}$ data shown here.

Method 4: the point where the error function in the modelled $C_i$ has reached 99\% of its maximum value (99\%ERF). 
This can be interpreted as the position where the error function can describe only~1\% of the remaining increase in $C_i$.
The choice of the 99\% threshold is arbitrary, and researchers implementing this method could choose any threshold. 

Method 5: Evaluate only the second term of equation \ref{eq:cumcount_equal} or \ref{eq:cumcount_unequal} with $X \to \infty$.
This method is only sensible if the fitted $C_i(X)$ curves are an excellent fit to the data around $X_0$ and above.
For the equal-concentration case (equation \ref{eq:cumcount_equal}), the interfacial excess from this method is found from:
\begin{equation}
  N_{i, \mathrm{XS}} = \lim_{X \to \infty} C_i(X)  = G W \sqrt{2 \pi} \,,
\end{equation}
with the uncertainty (excluding covariance between $G$ and $W$):
\begin{equation}
  \Delta N_{i, \mathrm{XS}} = N_{i, \mathrm{XS}} \sqrt{ \left(\frac{\Delta G}{G}\right)^2 + \left(\frac{\Delta W}{W}\right)^2} \,.
\end{equation}
$\Delta G$ and $\Delta W$ are the ESDs in $G$ and $W$, determined by least-squares fitting of equation \ref{eq:cumcount_equal} to the experimental data.
For the unequal-concentration case (equation \ref{eq:cumcount_unequal}),
\begin{equation}
  N_{i, \mathrm{XS}} = \lim_{X \to \infty} C_i(X) = \left(G - \frac{C_{i,2} - C_{i,1}}{2} \right) W \sqrt{2\pi} \,,
\end{equation}
with the uncertainty
\begin{equation}
  \Delta N_{i, \mathrm{XS}} = N_{i, \mathrm{XS}} \sqrt{ \left( \frac{\Delta G}{  G - \frac{C_2 - C_1}{2} }\right)^2 + \left(\frac{\Delta W}{W}\right)^2 + \frac{\Delta C_1^2 + \Delta C_2^2}{4\left(G - \frac{C_2 - C_1}{2} \right)^2}} \,,
\end{equation}
where $\Delta C_1$ and $\Delta C_2$ are the ESDs in the solute concentration in each grain, again determined from least-squares fitting of equation \ref{eq:cumcount_unequal}.
This method can loosely be interpreted as assigning $X_2 = \infty$.

For the Optimized ZIRLO reconstruction, the first four methods provide different $X_2$ but approximately equal and $N_{\XS}$, within the large uncertainties.
In this scenario, any of the first four methods are essentially equivalent.
In contrast, each method (except 2 and 3) gives different $N_{\ce{Mn},\XS}$ in the steel reconstruction due to the lower $\Delta C_i$ in this data.
For this data, a choice of method is important.
Method~5 gives comparable $N_{i,\XS}$ to the the other methods but far greater uncertainty because, qualitatively, it also considers the ESD in $W$.

\subsection{Effect of bin width}

\begin{figure}[b!]
  \centering
  \includegraphics[width=\linewidth]{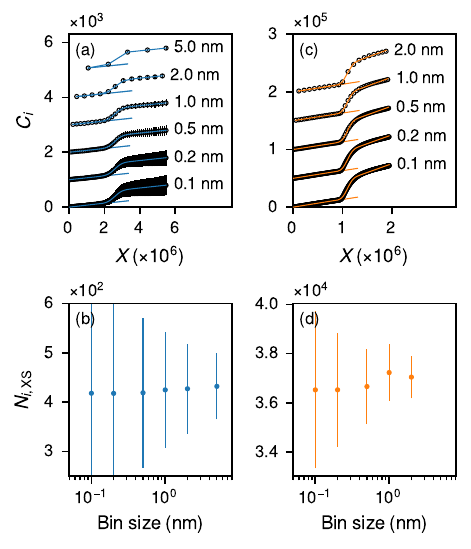}
  \caption{The effect of bin width on the interfacial excess calculated for (a,b) the Optimized ZIRLO reconstruction and (b,d) the steel reconstruction. (a,c) Cumulative ion count profiles with varying bin widths, indicated by the annotations. (b,d) Effect of bin width on $N_{\XS}$, calculated using method 4.}
  \label{fig:binsize}
\end{figure}

The results provided above show that the uncertainty is $\Delta N_{\XS}$ is dominated by compositional uncertainty.
One way to decrease this uncertainty in a composition profile is to increase the bin width.
Figure \ref{fig:binsize}(a,c) shows cumulative ion count profiles prepared with different bin widths, indicated by the annotations.
$N_{\XS}$ was calculated for each profile using method 4, as this was found to be most robust to varying step sizes.
These $N_{\XS}$ are shown in figure \ref{fig:binsize}(b,d).
There is little change in $N_{\XS}$ with bin size for either reconstruction, relative to the uncertainties.
Increasing the bin size decreases $\Delta N_{\XS}$ as this reduces $C_{i}$ at $X_2$.
For large step sizes there is a slight increase in the $\Delta C_{i,1}$ at $X_2$ associated with the reduced number of data points in grain 1, but this does not significantly impact $\Delta N_{\XS}$.
Overall, larger bin widths reduce relative uncertainty in $N_{\XS}$, although with diminishing returns.
It must be ensured that there are $>2$ data points in each grain so that linear regions in each grain can be accurately modelled.

\section{Discussion}

\begin{figure}[b!]
  \centering
  \includegraphics
  [width=\linewidth]
  {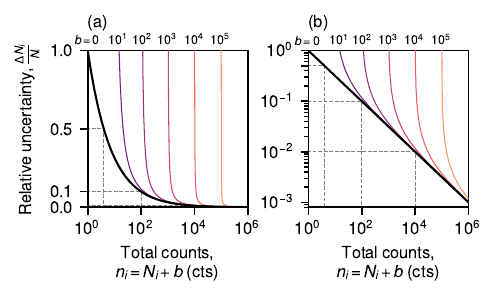}
  \caption{Relative uncertainty $\Delta N_i / N_i$ in terms of total peak counts $n_i = N_i + b$ on (a) semi-log and (b) log-log axes. Annotations at the top indicate the number of background counts $b$ for each line. Dashed lines indicate some required counts to reach 50\%, 10\% and 1\% relative uncertainty.}
  \label{fig:relerr}
\end{figure}

A key result from this analysis is that the compositional uncertainty dominates the total uncertainty in an interfacial excess calculation.
This is a fundamental and unavoidable limitation of compositional analysis using atom probe data \autocite{Gedcke_2009,Larson_2013}.
Figure \ref{fig:relerr} shows a plot of the relative uncertainty $\Delta N_i / N_i = \Delta c_i / c_i$ in terms of $N$, on (a) semi-log axes and (b) log-log axes.
The thick black line indicates the condition where background counts are zero.
Dashed lines indicate the $N_i$ required to achieve a certain relative uncertainties.
For example, $N_i = 4$ counts will have 50\% relative uncertainty and $N_i = 100$ will have 10\% relative uncertainty.
This point is illustrated in the Optimised ZIRLO data, which appears to be an ideal interfacial excess calculation: free of major evaporation rate fluctuations, a Gaussian concentration peak profile, and equal concentration in the grains on either side of the boundary.
Despite this, the relative uncertainty in $N_{\XS}$ is still $\approx$35\%.
This uncertainty is ultimately a result of the compositional uncertainty with position, and the only way to reduce this compositional uncertainty is to increase the sample size.
As shown, this can be done by increasing the bin width.
This may not always be possible, for example at the interface between a matrix and very small precipitates.
In this case, analysis will need to be performed for multiple similar precipitates and the results averaged, assuming the precipitates are equivalent.

The effect of the background has not been considered in the analysis so far.
If the counts are so low that the background level is appreciable, the counts must be considered as
\begin{equation}
  N_i = n_i - b \,,
\end{equation}
i.e. the total counts measured in that range ($n_i$) minus the counts from the background ($b$).
The uncertainty for this is \autocite{Larson_2013}
\begin{equation}
  \Delta N_i = \sqrt{n_i + b} = \sqrt{N_i + 2b} \,.
\end{equation}
This can substantially increase the required counts to obtain a certain relative uncertainty.
Figure \ref{fig:relerr} shows the effect of background counts on the relative uncertainty with different coloured lines for increasing $b$.
If the solute species mass spectrum peak has high background (for example, from a thermal tail of a peak with lower mass-to-charge) then more counts are needed to reach a certain relative uncertainty.
For example, if $N_i = 100$ counts also has $b = 100$ background counts then the relative uncertainty increases to $\approx 17$\%.

If method 1 is used to choose $X_2$, then equations \ref{eq:cumcount_equal} and \ref{eq:cumcount_unequal} do not need to be fit to the data.
Instead, equation \ref{eq:cumcount_linear} can be fit to grain 1, as only the uncertainty in $C_{i,0}$ is used for the uncertainty in $\Delta N_{\XS}$.
However, a choice must now be made over $X_1$, i.e., the end of the linear region and the start of the interfacial region. 
None of the methods described here can be used, as the relative uncertainty in grain~1 fluctuates greatly due to low counts (method 1) or the rely on the fitted error function (methods 2-4).
As such, the authors feel that such a procedure does not have any significant advantage over the procedures described in this work, except for its simplicity.

At the other extreme, method 5 requires equation \ref{eq:cumcount_equal} or \ref{eq:cumcount_unequal} perfectly describe the data $C_i(X)$ data.
If this condition is met then the $N_{i,\XS}$ produced is the most accurate.
However, it is the least precise, since it includes the uncertainty in the width parameter $W$ in calculating $\Delta N_{i,\XS}$.
For the most real experimental data sets, method 5 is the least practical.

\section{Conclusion}
Here, we derive general expressions for solute segregation to an interface and apply them to experimental atom probe data to determine the interfacial excess, $N_{i,\XS}$.
In contrast with many previous reports, we include the cumulative compositional uncertainty in the calculation of the interfacial excess.
We find that even when the segregation is obvious in the composition profile, the relative uncertainty in the interfacial excess can be high due to compositional uncertainty, ultimately arising from too few counts of the solute species in the ROI.
Increasing the number of ions in the bin (by increasing its width) somewhat improves the relative uncertainty, but this only useful for large data sets, where uncertainty is less likely to be an issue anyway.
We also compare some methods for defining the threshold between the interface and the matrix ($X_2$) and thus for calculating $N_{i,\XS}$.
When the compositional uncertainty is low, different methods for choosing the threshold can produce different values for the interfacial excess and a choice must be made by the researcher which method to use.
In this work, we found the most robust methods were the position of minimum relative uncertainty (method 1, MRU) and the point where the error function is 99\% complete (99\%ERF), though the disadvantage of the latter method is that it still requires specification of an arbitrary threshold.
When the relative uncertainty is high, the method for choosing the threshold value is less important, as the variation in interfacial excess counts between methods will most likely be less than the uncertainty from any method.

Based on these findings, the main recommendations by the authors for researchers performing an interfacial excess calculation are as follows:
\begin{itemize}
  \item Maximise the counts in the ROI used for the composition profile. This will minimise the uncertainty in $N_{i, \XS}$. In general, at first, this is done by maximising the cross-sectional area of the ROI.
  \item The uncertainty in the cumulative ion count is the sum of the uncertainties in the ion count at each previous step, i.e. equation \ref{eq:cumulativesum_uncertainty}.
  \item Ensure the ROI extends far beyond the interface by several hundred counts of the solute. This will minimise uncertainty in $C_0$, $C_1$ or $C_2$, and allow for all methods of choosing $X_2$.
  \item Set a large bin width such to minimise the uncertainty in $C_i$, but not so large that the composition profile and cumulative count profiles are not clearly visible or would rule-out your desired method for choosing $X_2$. Typical good choices will be 1--2 nm.
  \item Do not neglect the compositional uncertainty when making ``ladder plots'' or when calculating $N_\XS$, regardless of the method used.
\end{itemize}

\section*{Acknowledgements}

L. Tegg acknowledges Sima Aminorraya Yamini for providing the scientific problem which inspired this work.
The authors acknowledge Siyu Huang of the University of Sydney for collecting the Optimized ZIRLO data, and Hung-Wei (Homer) Yen of National Taiwan University for providing the microalloyed steel studied in this work.
The authors also acknowledge the technical and scientific support provided by Sydney Microscopy and Microanalysis at the University of Sydney, and the support of Microscopy Australia.



\printbibliography[heading=bibintoc,title=References]



\end{document}